\documentstyle[12pt]{article}
\def\journal{\topmargin .3in    \oddsidemargin .5in
        \headheight 0pt \headsep 0pt
        \textwidth 5.625in 
        \textheight 8.25in 
        \marginparwidth 1.5in
        \parindent 2em
        \parskip .5ex plus .1ex         \jot = 1.5ex}
\journal

\catcode`\@=11
\@addtoreset{equation}{section}

\def\theequation{\thesection.\arabic{equation}}

\newtoks\@stequation

\def\subequations{\refstepcounter{equation}%
  \edef\@savedequation{\the\c@equation}%
  \@stequation=\expandafter{\theequation}
  \edef\@savedtheequation{\the\@stequation}
  \edef\oldtheequation{\theequation}%
  \setcounter{equation}{0}%
  \def\theequation{\oldtheequation\alph{equation}}}

\def\endsubequations{\setcounter{equation}{\@savedequation}%
  \@stequation=\expandafter{\@savedtheequation}%
  \edef\theequation{\the\@stequation}\global\@ignoretrue
  \vspace*{-12pt} \\
  }

\def\marginnote#1{}
\catcode`\@=11
\def\section{\@startsection {section}{1}{0pt}{-3.5ex plus -1ex minus
 -.2ex}{2.3ex plus .2ex}{\raggedright\large\bf}}
\catcode`\@=12
\newskip\humongous \humongous=0pt plus 1000pt minus 1000pt

\newif\ifdtup

\def\Q{{\mathchoice
{\setbox0=\hbox{$\displaystyle\rm Q$}\hbox{\raise 0.15\ht0\hbox to0pt
{\kern0.4\wd0\vrule height0.8\ht0\hss}\box0}}
{\setbox0=\hbox{$\textstyle\rm Q$}\hbox{\raise 0.15\ht0\hbox to0pt
{\kern0.4\wd0\vrule height0.8\ht0\hss}\box0}}
{\setbox0=\hbox{$\scriptstyle\rm Q$}\hbox{\raise 0.15\ht0\hbox to0pt
{\kern0.4\wd0\vrule height0.7\ht0\hss}\box0}}
{\setbox0=\hbox{$\scriptscriptstyle\rm Q$}\hbox{\raise 0.15\ht0\hbox to0pt
{\kern0.4\wd0\vrule height0.7\ht0\hss}\box0}}}}
\def\C{{\mathchoice
{\setbox0=\hbox{$\displaystyle\rm C$}\hbox{\hbox to0pt
{\kern0.4\wd0\vrule height0.9\ht0\hss}\box0}}
{\setbox0=\hbox{$\textstyle\rm C$}\hbox{\hbox to0pt
{\kern0.4\wd0\vrule height0.9\ht0\hss}\box0}}
{\setbox0=\hbox{$\scriptstyle\rm C$}\hbox{\hbox to0pt
{\kern0.4\wd0\vrule height0.9\ht0\hss}\box0}}
{\setbox0=\hbox{$\scriptscriptstyle\rm C$}\hbox{\hbox to0pt
{\kern0.4\wd0\vrule height0.9\ht0\hss}\box0}}}}

\font\fivesans=cmss10 at 4.61pt
\font\sevensans=cmss10 at 6.81pt
\font\tensans=cmss10
\newfam\sansfam
\textfont\sansfam=\tensans\scriptfont\sansfam=\sevensans\scriptscriptfont
\sansfam=\fivesans
\def\sans{\fam\sansfam\tensans}
\def\Z{{\mathchoice
{\hbox{$\sans\textstyle Z\kern-0.4em Z$}}
{\hbox{$\sans\textstyle Z\kern-0.4em Z$}}
{\hbox{$\sans\scriptstyle Z\kern-0.3em Z$}}
{\hbox{$\sans\scriptscriptstyle Z\kern-0.2em Z$}}}}

\mathchardef\endbar="375

\def\ceilfill{$\raise3pt\hbox{$\mathsurround=0pt\mathord\endbar$}
  \mkern-2mu \xleaders\hbox{$\mkern-5mu
  \mathord-\mkern-5mu$}\hfill\mkern-7mu
  \raise3pt\hbox{$\mathsurround=0pt\mathord\endbar$}$}

\def\floorfill{$\raise9pt\hbox{$\mathsurround=0pt\mathord\endbar$}
  \mkern-2mu \xleaders\hbox{$\mkern-5mu
  \mathord-\mkern-5mu$}\hfill\mkern-7mu
  \raise9pt\hbox{$\mathsurround=0pt\mathord\endbar$}$}

\def\overcontract#1{\mathop{\vbox{\ialign{##\crcr\noalign{\kern3pt}
  \ceilfill\hskip6pt\crcr\noalign{\kern3pt\nointerlineskip}
  $\hfil\displaystyle{#1}\hfil$\crcr}}}}

\def\undercontract#1{\mathop{\vtop{\ialign{##\crcr
  $\hfil\displaystyle{#1}\hfil$\crcr\noalign{\kern3pt\nointerlineskip}
  \floorfill\hskip6pt\crcr\noalign{\kern3pt}}}}}
\def\a{\alpha}
\def\b{\beta}

\def\d{\delta}
\def\e{\epsilon}

\def\t{\tau}
\def\p{\pi}

\def\s{\sigma}

\def\o{\omega}

\def\et{\eta}

\def\D{\Delta}

\def\bB{\bar{B}}

\def\alphab{\bar{\alpha}}
\def\~{\tilde}
\def\tP{\tilde P}
\def\tc{\tilde c}
\def\hh{\tilde h}
\def\tL{\tilde{L}}
\def\tT{\tilde{T}}

\def\sL{{\cal L}}

\def\H{{\cal H}}

\def\T{{\cal T}}
\def\C{{\cal C}}
\def\cO{{\cal O}}

\def\hg{\hat{g}}
\def\bul{$\bullet \;\,$}
\def\half{{1\over2}}
\def\lrp{\stackrel{\leftrightarrow}{\partial}}
\def\lrbp{\stackrel{\leftrightarrow}{\bar\partial}}

\def\pa{\partial}

\def\ra{\rightarrow}

\def\Tr{{\rm Tr}}

\def\xx{\hbox{ }^*_*}

\def\nl{\newline}
\def\thefootnote{\fnsymbol{footnote}}

\begin{document}
\begin{titlepage}
 \noindent May 1995     \hfill        UCB-PTH-95/15 \\
 hep-th 9505125 \hfill     LBL-37255 \\
 \hfill

\centerline{\large \bf The Generic World-Sheet Action}
\centerline{\large \bf of Irrational Conformal Field Theory
\footnote{This work was
supported in part by the Director, Office of
Energy Research, Office of High Energy and Nuclear Physics, Division of
High Energy Physics of the U.S. Department of Energy under Contract
DE-AC03-76SF00098 and in part by the National Science Foundation under
grant PHY90-21139.}
\footnote{Invited talk by MBH
 at the conference ``Strings '95: Future Perspectives
in String Theory'', University
of Southern California, March 13-18.}}

\vspace*{0.4cm}
\centerline{\footnotesize K. Clubok\footnote{E-mail:
CLUBOK@PHYSICS.BERKELEY.EDU}~~and
M.B. Halpern\footnote{E-mail: MBHALPERN@LBL.GOV}}
\vspace*{0.1cm}
\centerline{\footnotesize \it Department of Physics, University of California}
\centerline{\footnotesize \it
Theoretical Physics Group, Lawrence Berkeley Laboratory}
\centerline{\footnotesize \it Berkeley, California 94720}
\centerline{\footnotesize \it USA}
\vspace*{0.9cm}

\begin{abstract}
We review developments in the
world-sheet action formulation of the generic irrational
conformal field theory, including the non-linear and the linearized
forms of the action.
These systems form a large class of spin-two gauged WZW actions which exhibit
exotic gravitational couplings.  Integrating out the gravitational
field, we also speculate on a connection with sigma models.
\end{abstract}
\end{titlepage}

\setcounter{footnote}{0}
\renewcommand{\thefootnote}{\alph{footnote}}

\section{Affine Lie Algebra and Conformal Field Theory}

Affine Lie algebra, or current algebra on the circle,
was discovered independently in mathematics  \cite{KM}
and physics \cite{BH}.
The affine algebras have played a central role in the construction of
new conformal field theories \cite{BH,H71,KZ,GKO,HK}, which have historically
been found first in the Hamiltonian or operator
 formulation and only later in their
corresponding action formulations \cite{Nov,Wit,GW,GW2,HY}.

One reason for this order of events is that Hamiltonians with different
local symmetries correspond to qualitatively different  actions,
 so that generalization to larger classes of actions is not
always straightforward.  For example, the affine-Sugawara
construction \cite{BH,H71,KZ}
is described by the WZW action \cite{Nov,Wit}, the coset
constructions \cite{BH,H71,GKO}
are described by the spin-one gauged WZW actions \cite{GW,GW2},
and the generic irrational conformal field
theory \cite{HK,nuc} is described by the spin-two gauged WZW
actions \cite{HY,bch}.

On the other hand, all these theories
 are uniformly included
in the general affine-Virasoro construction \cite{HK,Rus},
\begin{equation}
T=L^{ab}\xx J_aJ_b\xx\quad, \quad a,b=1\ldots\dim g
\end{equation}
which is quadratic in the currents $J_a$ of the general affine
algebra $\hg$.
The coefficient $L^{ab}$ is called the inverse inertia tensor, in
analogy with the spinning top.
In the general case, the inverse inertia tensor must
satisfy the Virasoro master equation \cite{HK,Rus}, whose solutions include
the affine-Sugawara and coset constructions, and a very large
number of new conformal field theories.

The generic affine-Virasoro construction has irrational central charge,
even on positive integer level of the compact affine algebras.  As
a consequence, the study of this class of theories
 is called irrational conformal field
theory (ICFT),
\begin{equation}
\mbox{ICFT} \supset\supset \mbox{RCFT}
\end{equation}
which includes rational conformal field theory (RCFT) as a small subspace.

The development of ICFT has moved through a number of stages, including:
{\samepage
\nl\bul exact unitary irrational solutions \cite{nuc} of the master equation
\nl\bul partial classification \cite{gt,ggt} of the solution space of the
master equation
\nl\bul generalized KZ equations on the sphere \cite{fc} and the
torus \cite{tor}
\nopagebreak \nl \nopagebreak \bul the generic world-sheet action
\cite{HY,bch} of ICFT.
\nl
See Ref.~[19] for a broad review of irrational conformal field theory.
}

In this talk, we  focus on the Hamiltonian and action formulation
of the generic ICFT on simple $g$.
In this formulation, the geometry of the action
is determined by the local symmetry group of the Hamiltonian.
The local symmetry of the generic
Hamiltonian is smaller than that of the coset constructions, and the
corresponding generic action is a large set of
spin-two gauged WZW models.  The spin-two gauge fields are
gravitational fields with exotic matter
couplings which generalize and include the coupling of the
conventional world-sheet metric.
We will discuss in particular the non-linear and linearized forms of the
action and a speculative connection with sigma models.

\section{Fundamentals of ICFT}

We begin by briefly reviewing some basic facts about ICFT which are
essential to understanding the action formulation.

For any Lie $g$, the general current algebra $\hat g$
is \cite{KM,BH}
\begin{equation}
J_a(z)J_b(w)={G_{ab}\over(z-w)^2} + if_{ab}{}^c {J_c(w)\over z-w}
 + \mbox{reg.}
\end{equation}
where $a,b=1\ldots\dim g$, and $f_{ab}{}^c$ and $G_{ab}$ are the
structure constants and generalized metric of $g$.  For the example of
simple compact $g$, one has $G_{ab}=k\et_{ab}$, where
$\et_{ab}$ is the Killing metric of $g$ and $k$ is the level of the
affine algebra.

On each level of $\hat g$, the
general affine-Virasoro
construction is \cite{HK,Rus}
{\samepage
\begin{subequations}
\begin{equation}
T(z)=L^{ab}\xx J_a(z)J_b(z)\xx
\label{genvir}
\end{equation}
\begin{equation}
L^{ab}=2L^{ac}G_{cd}L^{db}-L^{cd}L^{ef}{f_{ce}}^{a}
{f_{df}}^{b}-L^{cd}
{f_{ce}}^{f}{f_{df}}^{(a}L^{b)e}
\label{VME}
\end{equation}
\begin{equation}
c=2G_{ab}L^{ab}
\label{charge}
\end{equation}
\end{subequations}
}
where (\ref{VME}) is the Virasoro master equation.
The master equation is a large set of coupled quadratic equations for
the inverse inertia tensor $L^{ab}$.
For each solution $L^{ab}$ of the master equation, one obtains a
conformal field theory with central charge $c$ given in (\ref{charge}).
Generically-irrational central charge, even on positive integer levels
of the affine algebra, is an immediate consequence of the structure
of the master equation.
In what follows, we confine our remarks to simple compact $g$, which is
all that is needed below.

\noindent a) Affine-Sugawara construction \cite{BH,H71,KZ}.
The affine-Sugawara construction is the simplest solution
of the master equation, with
\begin{equation}
L^{ab}_g = {\et^{ab}\over 2k+Q_g}\quad, \quad c_g={2k\dim g \over 2k+Q_g}
\end{equation}
where $Q_g\et_{ab}=-f_{ac}{}^d f_{bd}{}^c$.

\noindent b) K-conjugation covariance \cite{BH,HK,H71}.
A central feature of the master equation is that its solutions come in
K-conjugate pairs $L$ and $\tL$, where
\begin{equation}
L^{ab}+\tL^{ab}=L^{ab}_g\quad, \quad c+\tc=c_g.
\end{equation}
The corresponding K-conjugate stress tensors $T$ and $\tT$,
\begin{equation}
T(z)\tT(w)=\mbox{reg.}\quad, \quad T_(z)+\tT(z)=T_g(z)
\end{equation}
commute and sum to the affine-Sugawara stress tensor $T_g$.

K-conjugation is used to generate additional solutions, as
in the
familiar case of the coset
constructions \cite{BH,H71,GKO} $T_h+T_{g/h}=T_g$.
At the level of dynamics, K-conjugation also provides the minimal
local symmetry of any ICFT, as discussed below.

\noindent c) Semi-classical solutions \cite{nuc,hl}.  On simple $g$, the
generic
solutions of the master equation live in level-families $L^{ab}(k)$ whose
high-level forms are
\begin{subequations}
\begin{equation}
L^{ab}={P^{ab} \over 2k} + \cO(k^{-2})\quad, \quad
\tL^{ab}={\tP^{ab} \over 2k} + \cO(k^{-2})
\end{equation}
\begin{equation}
L^{ab}+\tL^{ab}=L_g^{ab}={\et^{ab} \over 2k} + \cO(k^{-2})
\end{equation}
\begin{equation}
c = \mbox{rank } P + \cO(k^{-1})\quad, \quad
\tilde c= \mbox{rank } \tP + \cO(k^{-1})\quad, \quad
c_g=\dim g + \cO(k^{-1})
\label{highc}
\end{equation}
\begin{equation}
P^2=P \quad , \quad \tP^2=\tP \quad , \quad
P \tP =0 \quad, \quad P+\tP=1
\end{equation}
\label{high-level}
\end{subequations}
where $P$ and $\tP$ are the high-level projectors of the $L$ and the $\tL$
theory respectively.

The partial classification \cite{gt,ggt} of ICFT by graph theory is based on
these high-level forms, and it is believed \cite{Review} that the generic
level-family is generically unitary on positive integer levels
of $\hat g$.

\noindent d) Spin-two gauge theories \cite{HY,progress,bch}.
The basic Hamiltonian of the general affine-Virasoro construction is
\begin{equation}
H_0=L_0+\bar{L}_0 = L^{ab} (\xx J_aJ_b + \bar{J}_a\bar{J}_b \xx)_0
\end{equation}
where the barred currents $\bar J$ are right-mover copies of the
left-mover currents $J$.  For the $g/h$ coset constructions the
symmetry algebra of $H_0$ is affine $h\times h$, which leads to
a world-sheet description by spin-one (or Lie-algebra)
gauged WZW models \cite{GW,GW2}.

In the space of all conformal field theories,
the coset constructions are only special
points of higher symmetry.  For the generic affine-Virasoro
construction $L^{ab}$, the symmetry group of $H_0$ is
Diff S$_1$ $\times$ Diff S$_1$, generated by the left- and right-mover stress
tensors
\begin{equation}
\tT=\tL^{ab}\xx J_a J_b\xx \quad , \quad
\bar{\tT}=\tL^{ab}\xx\bar J_a\bar J_b\xx
\end{equation}
 of the K-conjugate theory
$\tL^{ab}$.  As a consequence, the world-sheet
action of the generic ICFT is a spin-two gauged WZW model, as discussed
in the following section.

\section{The Exotic Gravities of ICFT}
\vspace*{-12pt}
\subsection{The Generic Affine-Virasoro Hamiltonian}

The classical basic Hamiltonian of the generic level-family $L^{ab}(k)$ on
simple $g$ is
\begin{equation}
H_0=\int_0^{2\p} d\s \H_0\quad, \quad \H_0={1\over2\p}L^{ab}_\infty
(J_aJ_b + \bar J_a\bar J_b)
\label{basHam}
\end{equation}
where $L^{ab}_\infty=P^{ab}/2k$ is the high-level form of $L^{ab}$ in
Eq.~(\ref{high-level}). The classical currents $J_a, \bar J_a$ are
taken as Bowcock's canonical forms \cite{bow}, which satisfy the bracket
algebra of affine $g \times g$.
The coset constructions, with (bracket) affine symmetry, are included
in (\ref{basHam}) when $L=L_{g/h}=L_g-L_h$, but we consider only
the generic $L^{ab}$ for which,
as in the quantum theory, the
local symmetry algebra of $H_0$ is Diff S$_1$ $\times$ Diff S$_1$.
The generators of the diffeomorphism groups
are the conformal stress
tensors of the commuting K-conjugate theory,
\begin{equation}
\tilde{L}^{ab}_\infty J_a J_b\quad, \quad
 \tilde{L}^{ab}_\infty \bar{J}_a \bar{J}_b
\label{Stress2}
\end{equation}
where $\tL^{ab}_\infty=\tP^{ab}/2k$ is the high-level form of $\tL^{ab}$ in
Eq.~(\ref{high-level}).  Since these classical generators satisfy the
Virasoro algebra without central extension, they form a
set of first class constraints of $H_0$.

Following Dirac, one obtains the full Hamiltonian \cite{HY} of the generic
theory $L$,
{\samepage
\begin{subequations}
\begin{equation}
H=\int_0^{2\pi} d\sigma {\cal H}\quad, \quad
{\cal H}  =  {\cal H}_0 + v \cdot K(\tilde{L}_\infty)
\end{equation}
\begin{equation}
v \cdot K(\tilde{L}_\infty) = {1\over2\pi}\tilde{L}^{ab}_\infty (v J_a J_b
    + \bar{v} \bar{J}_a \bar{J}_b)
\end{equation}
\label{fullham}
\end{subequations}
}
where the K-conjugate stress tensors in $v\cdot K$ play the role of
Gauss' law and $v,\bar v$ are multipliers.  The multipliers form a
spin-two gauge field or gravitational field
 on the world-sheet, which is called the
 K-conjugate metric.
This Hamiltonian generalizes and includes the WZW Hamiltonian (which is
included when $P=1, \tP=0$) and the conventional world-sheet metric
formulation of the WZW model (which is included when $P=0, \tP=1$).

More generally, the K-conjugate metric is an exotic gravity
because it exhibits exotic, $\tL^{ab}$-dependent coupling only to the
``K-conjugate matter'', which is,
loosely speaking, only ``half'' the matter.

\subsection{The Non-Linear Action}

The action corresponding to $H$ in (\ref{fullham})
 is the non-linear form of the generic
affine-Virasoro action \cite{HY},
\begin{subequations}
\begin{equation}
     S  =  \int d\tau d\sigma ({\cal L} + \Gamma)
\end{equation}
\begin{eqnarray}
     {\cal L} & = & {1\over 8\pi}e_i{}^aG_{bc}\e_j{}^c\biggl[
        \left[f(Z)
          +\alpha\alphab\omega\tilde{P}\omega^{-1}f(Z)\tilde{P}\right]_a{}^b
           \left(\dot{x}^i\dot{x}^j - x^{\prime i}x^{\prime j}\right)
\nonumber \\
       & &+\alpha\left[f(Z)\tilde{P}\right]_a{}^b
             \left(\dot{x}^i\dot{x}^j  + x^{\prime i}x^{\prime j}
                         +\dot{x}^{(i}x^{j)\prime}\right)
\nonumber \\
      &  &+\alphab
               \left[\omega\tilde{P}\omega^{-1}f(Z)\right]_a{}^b
             \left(\dot{x}^i\dot{x}^j  + x^{\prime i}x^{\prime j}
                         -\dot{x}^{(i}x^{j)\prime}\right)
\nonumber \\
      & &+\left[1-f(Z)
          +\alpha\alphab\omega\tilde{P}\omega^{-1}f(Z)\tilde{P}\right]_a{}^b
             \left(\dot{x}^{[i}x^{j]\prime}\right)\biggr]
\label{badl}
\end{eqnarray}
\label{eqafvirl}
\begin{equation}
f(Z)  \equiv  [1-\alpha \alphab Z]^{-1}, \hskip 15pt
Z\equiv \tilde{P}\omega\tilde{P}\omega^{-1}, \hskip 15pt
\alpha \equiv {1-v \over 1+v}, \hskip 15 pt
\bar{\alpha} \equiv {1-\bar{v} \over 1+ \bar{v}}
\end{equation}
\end{subequations}
which is the world-sheet description of the generic theory $L$.
Here, $x^i$ and $e_i{}^a, i=1\ldots\dim g$
are the coordinates and left-invariant vielbein on the group manifold $G$.
Also, $\omega(g)_a{}^b$ is
the adjoint action of $g\in G$, and $\Gamma$ is the WZW term.
The non-linear action reduces to the WZW action when $\tP=0$, and to the
world-sheet metric formulation of the WZW model when $\tP=1$.

We call attention to the exotic, non-linear coupling of the
K-conjugate metric $(\a,\bar\a)$ in the action (\ref{eqafvirl}). In spite
of this non-linearity, the action exhibits
Lorentz, conformal and local Weyl symmetries, and
the expected world-sheet
diffeomorphism invariance.  The diffeomorphism group is called
Diff S$_2(K)$ because it is associated to the commuting
K-conjugate theory.

The K-conjugate metric can be written in standard form,
\begin{equation}
\tilde{h}_{mn}\equiv e^{-\phi}\pmatrix{ -v \bar{v} &
       {1 \over 2}(v- \bar{v}) \cr {1 \over 2}(v-\bar{v}) & 1 }, \hskip 10pt
\sqrt{-\tilde h} \tilde h^{mn} = {2 \over v+\bar v}
  \pmatrix{ -1 & {1\over2}(v-\bar v) \cr {1\over2}(v-\bar v) & v \bar v }
\label{tmetric}
\end{equation}
and $\tilde h_{mn}$
transforms under Diff S$_2(K)$ as a second-rank tensor field. The
Diff S$_2(K)$ transformations of the matter are given in Refs.~[11] and [13].
In the discussion below, we give the corresponding
matter transformations
for the linearized form of the action.

The gravitational stress tensor of the K-conjugate metric is defined in the
usual way,
\begin{equation}
\~\theta^{mn}={2\over \sqrt{-\tilde{h}}} {\delta S \over \delta
 \tilde{h}_{mn}}
\label{gravtens}
\end{equation}
and, in the conformal gauge,
\begin{equation}
v=\bar v=1\quad, \quad
\alpha=\bar\alpha=0\quad,\quad
\sqrt{-\tilde h}\tilde h^{mn}=\pmatrix{-1 & 0 \cr 0 & 1}
\end{equation}
this prescription reproduces the conformal stress tensor of the
$\tL$ theory, as it should. It follows that $\tilde h_{mn}$ is the
world-sheet metric of the $\tL$ theory.

The chiral currents of the underlying affine-Virasoro construction
(\ref{genvir}) are also found \cite{HY} in the conformal gauge.
In other gauges, the currents are gauge-equivalent to chiral
currents.

\subsection{The Linearized Action}

In the equivalent linearized form \cite{bch}, the generic action
is clearly seen as a
large class of
spin-two gauged WZW models,
\begin{subequations}
\begin{equation}
S'=S_{WZW}+\int d^2z \Delta \sL_B
\end{equation}
\begin{eqnarray}
\D \sL_B &=& {\a\over\p y^2} \tL^{ab}_{\infty}\Tr(\T_aB)\Tr(\T_bB)
\nonumber \\
 & & + {\bar\a\over\p y^2} \tL^{ab}_\infty\Tr(\T_a\bB)\Tr(\T_b\bB)
\nonumber \\
 & & - {1\over\p y} \Tr(\bar D_BgD_Bg^{-1})
\end{eqnarray}
\begin{equation}
D_B\equiv \pa+iB\quad, \quad \bar D_B\equiv\bar\pa+i\bB.
\label{covar}
\end{equation}
\label{maction1}
\end{subequations}
Here, $g(\T)$ is the group element in irrep $\T$ of $g$,
$S_{WZW}$ is the Wess-Zumino-Witten action and $y\sim k^{-1}$ is
associated to the trace normalization of $\T$.
The quantities $B=B^a\T_a, \bar B=\bar B^a\T_a$ are a set of auxiliary
fields,  called the connections for reasons which will
be clear below.
Integration of the connections gives the non-linear form
(\ref{eqafvirl}) of the action.

The action (\ref{maction1}) describes the generic theory $L$ as a
spin-two gauging of the WZW action by the K-conjugate theory $\tL$.
The couplings of the gauge field $\hh_{mn}(\a,\bar\a)$ are quite
simple in this form.
The linearized action is invariant under the
Diff S$_2(K)$ transformations
\begin{subequations}
\begin{equation}
\delta \alpha   =  -\bar\partial\xi+\xi\lrp\alpha, \hskip 15pt
\delta \bar\alpha = -\partial\bar\xi + \bar\xi\lrbp\bar\alpha
\label{alpha}
\end{equation}
\begin{equation}
\delta g=gi\lambda-i\bar\lambda g
\label{gtrans}
\end{equation}
\begin{equation}
\delta B=\partial\lambda+i[B,\lambda], \hskip 10pt
\delta \bar B=\bar\partial\bar\lambda+i[\bar B,\bar\lambda]
\label{Btrans}
\end{equation}
\begin{equation}
\lambda\equiv\lambda^a\T_a,\hskip 15pt \bar\lambda\equiv\bar\lambda^a\T_a
\end{equation}
\begin{equation}
\lambda^a  \equiv  2\xi\tilde L^{ab}_\infty B_b, \hskip 10 pt
\bar\lambda^a \equiv 2\bar\xi\tilde L^{ab}_\infty \bar B_b
\label{lambda}
\end{equation}
\label{trans1}
\end{subequations}
where $\xi,\bar\xi$ are the diffeomorphism parameters.
The transformation of the K-conjugate metric in (\ref{alpha}) is
the usual transformation of a second-rank tensor
field, as noted above,
but the Diff S$_2(K)$ transformation of $B, \bar B$ and the matter is quite
exotic.

In particular, Eqs.~(\ref{gtrans},c) show that Diff S$_2(K)$ is locally
embedded in local Lie $g$ $\times$ Lie $g$, with the matter $g(\T)\in G$
and the connections
$B,\bar B$ transforming under local Lie $g$ $\times$ Lie $g$ as the
group element and the Lie $g$ $\times$ Lie $g$ connection respectively.
These transformation properties account for the covariant derivatives
in (\ref{covar}) and the
intriguing
resemblance of the linearized action (\ref{maction1}) to the usual
(Lie algebra) gauged WZW model.  Further discussion of the local embedding
and the
linearized action is given in Refs. [11], [13], and [19].

\section{Two world-sheet metrics}

We have seen that the world-sheet action of the generic theory
$L$ is a spin-two gauge theory, where the gauge field $\hh_{mn}$ is
the world-sheet metric of the $\tL$ theory.
Because  $\hh_{mn}$ couples only to the $\tL$ matter, it is
also possible \cite{HY} to introduce another spin-two gauge field
$h_{mn}$,
\begin{equation}
h_{mn}\equiv e^{-\chi}\pmatrix{ -u \bar{u} &
       {1 \over 2}(u- \bar{u}) \cr {1 \over 2}(u-\bar{u}) & 1 }, \hskip 10pt
\sqrt{-h} h^{mn} = {2 \over u+\bar u}
  \pmatrix{ -1 & \half(u-\bar u) \cr \half(u-\bar u) & u \bar u }
\end{equation}
which is the world-sheet metric of the $L$ theory.
This results in the {\em doubly-gauged action} $S_D$, with a
K-conjugate pair of world-sheet metrics $\hh_{mn}$ and $h_{mn}$.

The linearized form of the doubly-gauged action \cite{bch} is surprisingly
simple,
\begin{subequations}
\begin{equation}
S_D=S_{WZW}+\int d^2z \Delta \sL_D
\end{equation}
\begin{eqnarray}
\D \sL_D &=& {1\over\p y^2} (\a\tL^{ab}_{\infty}+\b L^{ab}_\infty)
\Tr(\T_aB)\Tr(\T_bB)
\nonumber \\
 & & + {1\over\p y^2} (\bar\a\tL^{ab}_\infty+\bar\b L^{ab}_\infty)
\Tr(\T_a\bB)\Tr(\T_b\bB)
\nonumber \\
 & & - {1\over\p y} \Tr(\bar D_BgD_Bg^{-1})
\end{eqnarray}
\begin{equation}
\a={1-v\over1+v}\quad,\quad \bar\a={1-\bar v\over 1-\bar v} \quad,\quad
\b={1-u\over1+u}\quad,\quad \bar\b={1-\bar u\over 1+\bar u}
\end{equation}
\label{doubS}
\end{subequations}
and may in fact
 be obtained from the action (\ref{maction1}) by the substitution
\begin{equation}
\alpha\tilde L^{ab}_\infty \rightarrow \alpha\tilde L^{ab}_\infty
                                      +\beta L^{ab}_\infty, \hskip 15pt
\bar\alpha\tilde L^{ab}_\infty \rightarrow \bar\alpha\tilde L^{ab}_\infty
                                      +\bar\beta L^{ab}_\infty.
\label{doubsub}
\end{equation}
Other forms of the doubly-gauged action, including the non-linear form,
are given in Refs.~[13] and [19].

The doubly-gauged action (\ref{doubS}) is invariant under two
commuting diffeomorphism groups \cite{bch}, called Diff S$_2(T)\times$
Diff S$_2(K)$, which are associated to the two world-sheet
metrics $h_{mn}$ and $\hh_{mn}$.
In particular, $h_{mn}$ is a
rank-two tensor under  Diff S$_2(T)$ but inert under Diff S$_2(K)$ and
vice versa for  $\tilde h_{mn}$.
The Diff S$_2(T)$ $\times$ Diff S$_2(K)$ transformations of the matter
and the connections are given in Ref.~[13]

In the doubly-gauged action (\ref{doubS}), the K-conjugate metrics
$\hh_{mn}(\a,\bar\a)$ and $h_{mn}(\b,\bar\b)$ are on equal footing.
When considering the $L$ theory, however, only $\hh_{mn}$ is dynamical, while
$h_{mn}$ provides a gravitational probe for the stress tensor of the
$L$ theory,
\begin{equation}
\theta^{mn}={2\over\sqrt{-h}} {\d S_D\over\d h_{mn}}
\label{stressh}
\end{equation}
in parallel with the stress tensor $\tilde\theta^{mn}$ of the
$\tL$ theory in (\ref{gravtens}).
In what follows, we refer to $\hh_{mn}$ and $h_{mn}$ as the $\tL$-metric
and the $L$-metric respectively.

\section{A Connection with Sigma Models}

An important open problem in the action formulation of ICFT is the possible
connection with sigma models.  In this section we sketch a speculative,
essentially classical derivation \cite{Review} of such a connection,
with surprising results.
The details of this derivation cannot be taken seriously until
one-loop effects are properly included.

One begins with the doubly-gauged action (\ref{doubS}) for the $L$
theory, and integrates the dynamical gauge fields $\a,\bar\a$ of
the $\tL$ metric.  This gives the $\d$-function constraints
on the connections,
\begin{equation}
\tP^{ab}B_a B_b=\tP^{ab}\bar B_a \bar B_b=0
\end{equation}
which are solved by $B_a=P_a{}^b b_b, \bar B_a=P_a{}^b \bar b_b$
with unconstrained $b,\bar b$. Then, one may integrate $b,\bar b$
to obtain the action
\begin{subequations}
\begin{equation}
S_{\b,\bar\b}=\int d\tau d\s(\sL+\Gamma)
\end{equation}
\begin{equation}
\sL=-{1 \over 8\p}G_{ab}(e_\tau{}^a e_\tau{}^b-e_\s{}^a e_\s{}^b)
  -{1\over 8\p} E^A(C^{-1})_A{}^B E_B
\end{equation}
\begin{equation}
E^A=\left(\matrix{
(e_\tau{}^a-e_\s{}^a), & (\bar e_\tau{}^a+\bar e_\s{}^a)}\right)\quad,
\quad E_B= \left( \matrix{
G_{bc}(e_\tau{}^c-e_\s{}^c) \cr G_{bc}(\bar e_\tau{}^c + \bar e_\s{}^c)}\right)
\end{equation}
\begin{equation}
C^{-1}=\left( \matrix{-\bar\b PM(\o)P & P\o PM(\o^{-1})P \cr
             P\o^{-1}PM(\o)P & \b PM(\o^{-1})P \cr } \right)
\end{equation}
\begin{equation}
M(\o)\equiv \left( (1+P(\o-1)P)(1+P(\o^{-1}-1)P)-\b\bar\b P\right)^{-1}
\end{equation}
\label{Sbb}
\end{subequations}
which is the conformal field theory of $L$ coupled to its world-sheet
metric $h_{mn}(\b,\bar\b)$.  Here, $e_\t{}^a=e_i{}^a\dot x^i,
e_\s{}^a =e_i{}^a x^{\prime i}$, and the barred quantities are defined
similarly with the right-invariant vielbeins $\bar e_i{}^a$.

Using the action (\ref{Sbb}) and
 the prescription (\ref{stressh}), one computes the stress
tensor of the $L$ theory,
\begin{subequations}
\begin{eqnarray}
\theta_{00}=\theta_{11}&=&{1\over16\pi}(e_\tau{}^a-e_\s{}^a)
(e_\tau{}^b-e_\s{}^b)
 G_{bc}(PF(\o)P)_a{}^c
\nonumber \\ & + &
 {1\over16\pi}(\bar e_\tau{}^a+\bar e_\s{}^a)(\bar e_\tau{}^b+\bar e_\s{}^b)
 G_{bc}(PF(\o^{-1})P)_a{}^c
\nonumber \\
\theta_{01}=\theta_{10}&=&{1\over16\pi}(e_\tau{}^a-e_\s{}^a)
(e_\tau{}^b-e_\s{}^b)
 G_{bc}(PF(\o)P)_a{}^c
\nonumber \\ & - &
 {1\over16\pi}(\bar e_\tau{}^a+\bar e_\s{}^a)(\bar e_\tau{}^b+\bar e_\s{}^b)
 G_{bc}(PF(\o^{-1})P)_a{}^c
\end{eqnarray}
\begin{equation}
F(\o)\equiv ((1+P(\o-1)P)(1+P(\o^{-1}-1)P))^{-1}
\end{equation}
\label{sigstress}
\end{subequations}
in the conformal gauge ($\b=\bar\b=0$) of the $L$ metric.  The matter degrees
of freedom in (\ref{sigstress})
are entirely
projected onto the $P$ subspace. As a consequence,
 one finds that the stress tensor is conformal
at high level with high-level central charge
\begin{equation}
c(L)=\mbox{rank } P +\cO(k^{-1})
\end{equation}
as it should be for the $L$ theory (see Eq.~(\ref{highc})).

Finally, one obtains the conformal field theory of $L$ as the
effective sigma model,
{\samepage
\begin{subequations}
\begin{equation}
S_{\rm eff}=
\int d\tau d\s \left[{1\over8\p} G_{ij}^{\rm eff}(\dot x^i \dot x^j-x'^i x'^j)
+{1\over4\p y} B_{ij}^{\rm eff} \dot x^i x'^j \right]
\end{equation}
\begin{equation}
G_{ij}^{\rm eff}=e_i{}^a e_j{}^b G_{cb} (N+N^T-1)_a{}^c\quad, \quad
B_{ij}^{\rm eff}=B_{ij}-y e_i{}^a e_j{}^b G_{cb} (N-N^T)_a{}^c
\end{equation}
\begin{equation}
N=\o P(1+P(\o-1)P)^{-1}P\quad,\quad
N^T=P(1+P(\o^{-1}-1)P)^{-1}P\o^{-1}
\end{equation}
\label{Sigma}
\end{subequations}
}
by evaluating the action (\ref{Sbb}) in the conformal gauge of the $L$ metric.
Here $B_{ij}$ is the WZW antisymmetric tensor field, and
$G_{ij}^{\rm eff}, B_{ij}^{\rm eff}$ are the space-time metric
and antisymmetric tensor field on the target space.  This sigma model
reduces to the WZW model when $P=1$, as it should.

The result (\ref{Sigma}) is an ordinary sigma model, but
the derivation above
illustrates the remarkable
fact that the question of conformal invariance of a sigma model depends
on the choice of the world-sheet metric (the gravitational coupling)
and its associated stress tensor.

The correct stress tensor (\ref{sigstress}) of the $L$ theory
 followed from the exotic coupling of the $L$ metric to the $L$
matter in the action (\ref{Sbb}), but one may
also consider the same sigma model (\ref{Sigma}) with the
distinct, conventional gravitational coupling \cite{cal},
\begin{equation}
\dot x^i \dot x^j - x'^i x'^j \ra \sqrt{-g_C}
g_C^{mn}\pa_m x^i\pa_n x^j
\end{equation}
of the conventional world-sheet metric $g_{mn}^C$, which gives the
conventional stress tensor $\theta_{mn}^C$,
\begin{equation}
\theta_{mn}^C={2\over\sqrt{-g_C}}{\d S\over \d g_{mn}^C}.
\end{equation}
It is unlikely that $\theta_{mn}^C$ is conformal in this case,
but the
question is not directly relevant because, as we have seen,
 $g^C_{mn}$ and $\theta_{mn}^C$ are
 not the world-sheet
metric and stress tensor of the $L$ theory.

We finally mention that Tseytlin \cite{ts3} and Bardak\c ci \cite{b}
 have studied a similar sigma model, which
is related to the bosonization of a generalized Thirring model \cite{kps}.
When $Q$ in
Tseytlin's (3.1) is taken as twice the high-$k$ projector $P$, his action
and the sigma model (\ref{Sigma})
are the same except for a dilaton term and
an overall minus sign for the kinetic term.
Further investigation will determine whether this intriguing
circumstance is more than a coincidence.

\section*{Acknowledgements}

We thank K. Bardak\c ci, C. Bachas, I. Bars,
J. deBoer, E. Kiritsis, N. Obers, and
A. Tseytlin for helpful discussions.
MBH also thanks the USC theory group for hosting this conference and
for the opportunity to present this
work.

This work was supported in part by the Director, Office
of Energy Research, Office of High Energy and Nuclear Physics, Division of
High Energy Physics of the U.S. Department of Energy under Contract
DE-AC03-76SF00098 and in part by the National Science Foundation under
grant PHY90-21139.

\end{document}